\documentclass[12pt]{article}
\usepackage{a4wide,epsfig,amsmath,amssymb,cite,scalefnt,graphicx}
\usepackage{slashed}
\usepackage{axodraw4j}
\usepackage{pstricks}
\usepackage{color}
\usepackage{changepage}
\usepackage{tabularx}
\usepackage{graphics}
\usepackage{array}

\begin{document}

\newcommand\drtj[1]{{\bf{Tim: #1}}} 

\def \inparg{\leftskip = 40pt\rightskip = 40pt}
\def \outparg{\leftskip = 0 pt\rightskip = 0pt}
\def\bolone{{\bf 1}}
\def\Ocal{{\cal O}}
\def\NSVZ{{\rm NSVZ}}
\def\DR{{\rm DRED}}
\def\DG{{\rm DREG}}
\def\DREDp{{\rm DRED}'}
\def\npb{{\it{Nucl.\ Phys.}\ }{\bf B}}
\def\plb{{{\it Phys.\ Lett.}\ }{ \bf B}}
\def\Dcal{{\cal D}}
\def\jhep{{\it JHEP}\ }
\def\be{\begin{equation}}
\def\ee{\end{equation}}
\def\bea{\begin{align}}
\def\eea{\end{align}}
\def\nn{\nonumber\\}
\def\n{\nonumber}
\def\tr{\hbox{tr}}
\def\barh{\overline h}
\def\prd{{{\it Phys.\ Rev.}\ }{\bf D}}

\def\prl{Phys.\ Rev.\ Lett.\ }
\def\zpc{Z.\ Phys.\ {\bf C}}
\def\msbar{{\overline{\rm MS}}}
\def\drbar{{\overline{\rm DR}}}

\def\etil{\tilde e}
\def\tlambda{\tilde \lambda}
\def\ttheta{\tilde \theta}
\def\tTheta{\tilde \Theta}
\def\tLambda{\tilde \Lambda}
\def\yhat{\hat y}
\def\that{\hat t}
\def\Gtil{\hat G}
\def\tZ{\tilde Z}
\def\tz{\tilde z}
\def\ov #1{\overline{#1}}
\def\thetabar{{\overline\theta}}
\def\alphadot{\dot\alpha}
\def\betadot{\dot\beta}
\def\Qbar{\overline Q}
\def\cbar{\overline c}
\def\fbar{\overline f}
\def\Fbar{\overline F}
\def\Lambdabar{\overline \Lambda}
\def\psibar{\overline{\psi}}
\def\betabar{\overline{\beta}}
\def\wt #1{\widetilde{#1}}
\def\Tr{\mathop{\rm Tr}}
\def\det{\mathop{\rm det}}
\def\wtl{\widetilde{\lambda}}
\def\wh{\widehat}
\def\th{{\theta}}
\def\bth{{\overline{\theta}}}
\def\eps{\epsilon}
\def\frak#1#2{{\textstyle{{#1}\over{#2}}}}
\def\nhalf{${\cal N} = \frak{1}{2}$}
\def\none{${\cal N} = 1$}
\def\ntwo{${\cal N} = 2$}

\def\ybar{\overline y}
\def\mbar{\overline m}
\def\tautil{\tilde\tau}
\def\chibar{\overline\chi}
\def\nutil{\tilde\nu}
\def\mutil{\tilde\mu}
\def\rhotil{\tilde\rho}
\def\sigtil{\tilde\sigma}
\def\gatil{\tilde\ga}
\def\Btil{\tilde B}
\def\Bbar{\overline B}
\def\Ttil{\tilde T}
\def\fhat{\hat f}
\def\Ahat{\hat A}
\def\Chat{\hat C}
\def\mbar{\overline{m}}
\def\mubar{\overline{\mu}}
\def\deltabar{\overline\delta}
\def\alphabar{\overline\alpha}
\def\half{{\textstyle{{1}\over{2}}}}
\def\frak#1#2{{\textstyle{{#1}\over{#2}}}}
\def\frakk[#1#2{{{#1}\over{#2}}}
\def\go{\rightarrow}
\def\lambdabar{\overline\lambda}
\def\lambdahatbar{\overline{\hat\lambda}}
\def\lambdahat{\hat\lambda}
\def\Dtil{\tilde D}
\def\Dhat{\hat D}
\def\Dbar{\overline D}
\def\Ftil{\tilde F}
\def\Fhat{\hat F}
\def\Atil{\tilde A}
\def\sigmabar{\overline\sigma}
\def\phibar{\bar\phi}
\def\phitilbar{\overline{\tilde{\phi}}}
\def\psitilbar{\overline{\tilde{\psi}}}
\def\Phibar{\overline\Phi}
\def\psibar{\overline\psi}
\def\Fbar{\overline F}
\def\TeV{{\rm TeV}}
\def\GeV{{\rm GeV}}
\def\Ghat{\hat\Gamma}
\def\Btil{\tilde B}
\def\Dtil{\tilde D}
\def\Etil{\tilde E}
\def\Ttil{\tilde T}
\def\Ytil{\tilde Y}
\def\Qtil{\tilde Q}
\def\Ltil{\tilde L}
\def\atil{\tilde \alpha}
\def\ctil{\tilde c}
\def\dtil{\tilde \delta}
\def\et{\tilde \epsilon}
\def\gtil{\tilde \gamma}
\def\mtil{\tilde \mu}
\def\ntil{\tilde \nu}
\def\rtil{\tilde \rho}
\def\stil{\tilde \sigma}
\def\xtil{\tilde \xi}
\def\ztil{\tilde \zeta}
\def\ttil{\tilde t}
\def\util{\tilde u}
\def\phitil{\tilde\phi}
\def\psitil{\tilde\psi}
\def\Ncal{{\cal N}}
\def\Ftil{\tilde F}
\def\Ytil{\tilde Y}
\def\alphadot{\dot\alpha}
\def\betadot{\dot\beta}
\def\deltadot{\dot\delta}
\def\Vhat{\hat V}
\def\Rhat{\hat R}
\def\Abar{\overline A}
\def\Bbar{\overline B}
\def\Mbar{\overline M}
\def\Nbar{\overline N}
\def\Hbar{\overline H}
\def\ahat{\hat a}
\def\bhat{\hat b}
\def\sy{supersymmetry}
\def\sic{supersymmetric}
\def\pa{\partial}
\def\pabar{\overline\partial}
\def\smgroup{$SU_3\otimes\ SU_2\otimes\ U_1$}
\def\stw{\sin^2\th_W}

\input amssym.def
\input amssym
\baselineskip 14pt
\parskip 6pt

\def\npb{Nucl. Phys. B}
\def\plb{Phys. Lett. B}
\def\pa{\partial}
\def\be{\begin{equation}}
\def\ee{\end{equation}}
\def\bea{\begin{align}}
\def\eea{\end{align}}
\def\nn{\nonumber\\}
\def\n{\nonumber}
\def\tr{\hbox{tr}}
\def\trhat{\hat\tr}
\def\barh{\overline h}
\def \de{\delta}
\def \De{\Delta}
\def \si{\sigma}
\def \Ga{\Gamma}
\def \ga{\gamma}
\def \nab{\nabla}
\def \pr{\partial}
\def \d{{\rm d}}
\def \tr{{\rm tr }}
\def \ta{{\tilde a}}
\def \hs{\hat s}
\def \hr{\hat r}
\def \br{\bar r}
\def \hG{{\hat G}}
\def \bI{{\bar I}}
\def \bL{{\bar L}}
\def \bT{{\bar T}}
\def \bO{{\bar O}}
\def \by{{\bar y}}
\def\bx{{\bar x}}
\def\ba{{\bar a}}
\def\bb{{\bar b}}
\def\bh{{\bar h}}
\def\bm{{\bar m}}
\def \bY{{\bar Y}}
\def \bW{{\bar W}}
\def \bQ{{\bar Q}}
\def \bl{{\lambda}}
\def \bpsi{{\bar \psi}}
\def \bsi{{\bar \sigma}}
\def \bchi{{\bar \chi}}
\def \bphi{{\bar \phi}}
\def \bta{{\bar ga}}
\def\ba{{\bar a}}
\def\bb{{\bar b}}
\def \rO{{\rm O}}
\def \l{\big \langle}
\def \r{\big \rangle}
\def \ep{\epsilon}
\def \vep{\varepsilon}
\def \half{{\textstyle {1 \over 2}}}
\def \thir{{\textstyle {1 \over 3}}}
\def \quar{{\textstyle {1 \over 4}}}
\def \ts{\textstyle}
\def\del{{\rm d}}
\def \A{{\cal A}}
\def \B{{\cal B}}
\def \C{{\cal C}}
\def \D{{\cal D}}
\def \E{{\cal E}}
\def \F{{\cal F}}
\def \G{{\cal G}}
\def \H{{\cal H}}
\def \I{{\cal I}}
\def \J{{\cal J}}
\def \K{{\cal K}}
\def \L{{\cal L}}
\def \M{{\cal M}}
\def \N{{\cal N}}
\def \O{{\cal O}}
\def \P{{\cal P}}
\def \Q{{\cal Q}}
\def \R{{\cal R}}
\def \S{{\cal S}}
\def \T{{\cal T}}
\def \V{{\cal V}}
\def \X{{\cal X}}
\def \Y{{\cal Y}}
\def \Z{{\cal Z}}
\def\fd{{\frak d}}
\def\fe{{\frak e}}
\def\fN{{\frak r}}
\def \tb{{\tilde {\rm b}}}
\def \tx{{\tilde {\rm x}}}
\def \ty{{\tilde {\rm y}}}
\def \tK{{\tilde K}}
\def \tsi{{\tilde \sigma}}
\def \h{{\rm h}}
\def \a{{\rm a}}
\def \b{{\rm b}}
\def \d{{\rm d}}
\def \e{{\rm e}}
\def \x{{\rm x}}
\def \y{{\rm y}}
\def\uF{\bar{F}}
\def\uG{\bar{G}}
\def\uA{\bar{A}}
\def\uB{\bar{B}}
\def\uC{\bar{C}}
\def\uD{\bar{D}}
\def\uE{\bar{E}}
\def\uH{\bar{H}}
\def\ual{\bar{\alpha}}
\def\ube{\bar{\beta}}
\def\uga{\bar{\gamma}}
\def\ude{\bar{\delta}}
\def\uet{\bar{ga}}
\def\uep{\bar{\epsilon}}
\def\hDel{\hat \Delta}
\def\wL{{\widetilde \L}}
\def\hrho{{\tilde \rho}}
\def\bxi{{\bar \xi}}
\def\blam{{\bar \lambda}}
\def \vphi{{\varphi}}
\def \tD{{\tilde D}}
\def\hbet{\beta}
\def\hhbet{{\hat \beta}}
\def\hB{{\hat B}}
\def\btil{\tilde \beta}
\def\cirk{\,{\raise1pt \hbox{${\scriptscriptstyle \circ}$}}\,}
\def\limu#1{\mathrel{\mathop{\sim}\limits_{\scriptscriptstyle{#1}}}}
\def\toinf#1{\mathrel{\mathop{\longrightarrow}\limits_{\scriptstyle{#1}}}}
\def \olr{{\raise6.5pt\hbox{$\leftrightarrow  \! \! \! \! \!$}}}

\font\ninerm=cmr9 \font\ninesy=cmsy9
\font\eightrm=cmr8 \font\sixrm=cmr6
\font\eighti=cmmi8 \font\sixi=cmmi6
\font\eightsy=cmsy8 \font\sixsy=cmsy6
\font\eightbf=cmbx8 \font\sixbf=cmbx6
\font\eightit=cmti8
\def\eightpoint{\def\rm{\fam0\eightrm}
  \textfont0=\eightrm \scriptfont0=\sixrm \scriptscriptfont0=\fiverm
  \textfont1=\eighti  \scriptfont1=\sixi  \scriptscriptfont1=\fivei
  \textfont2=\eightsy \scriptfont2=\sixsy \scriptscriptfont2=\fivesy
  \textfont3=\tenex   \scriptfont3=\tenex \scriptscriptfont3=\tenex
  \textfont\itfam=\eightit  \def\it{\fam\itfam\eightit}%
  \textfont\bffam=\eightbf  \scriptfont\bffam=\sixbf
  \scriptscriptfont\bffam=\fivebf  \def\bf{\fam\bffam\eightbf}%
  \normalbaselineskip=9pt
  \setbox\strutbox=\hbox{\vrule height7pt depth2pt width0pt}%
  \let\big=\eightbig  \normalbaselines\rm}
\catcode`@=11 %
\def\eightbig#1{{\hbox{$\textfont0=\ninerm\textfont2=\ninesy
  \left#1\vbox to6.5pt{}\right.\n@@space$}}}
\def\vfootnote#1{\insert\footins\bgroup\eightpoint
  \interlinepenalty=\interfootnotelinepenalty
  \splittopskip=\ht\strutbox %
  \splitmaxdepth=\dp\strutbox %
  \leftskip=0pt \rightskip=0pt \spaceskip=0pt \xspaceskip=0pt
  \textindent{#1}\footstrut\futurelet\next\fo@t}
\catcode`@=12 %
\def\today{\number\day\ \ifcase\month\or January\or February\or March\or
April\or May\or June\or July\or
August\or September\or October\or November\or December\fi, \number\year}
\def\now{\number\time}
\font \bigbf=cmbx10 scaled \magstep1

\input epsf

\numberwithin{equation}{section}

\begin{titlepage}
\begin{flushright}
LTH1043\\

\end{flushright}
\date{}
\vspace*{3mm}

\begin{center}
{\Huge Gradient flows in three dimensions}\\[12mm]
{\bf I.~Jack\footnote{{\tt dij@liv.ac.uk}}, D.R.T.~Jones\footnote{{\tt drtj@liv.ac.uk}} and
C.~Poole\footnote{{\tt c.poole@liv.ac.uk}}}\\

\vspace{5mm}
Dept. of Mathematical Sciences,
University of Liverpool, Liverpool L69 3BX, UK\\

\end{center}

\vspace{3mm}
\begin{abstract}
The $a$-function is a proposed quantity defined for quantum field theories
which
has a monotonic behaviour along renormalisation group flows, being related to
the $\beta$-functions via a gradient flow equation involving a positive
definite metric. We demonstrate the existence of a candidate $a$-function for renormalisable Chern-Simons theories in three dimensions, involving scalar and fermion fields, 
in both non-supersymmetric and supersymmetric cases.
\end{abstract}

\vfill

\end{titlepage}


\section{Introduction}

It is natural to regard quantum field theories as points on a manifold with
the couplings $\{g^I\}$ as co-ordinates, and with a natural flow
determined by the $\beta$-functions $\beta^I(g)$. At fixed points the quantum
field theory is scale-invariant and is expected to become a conformal field
theory. It was suggested by Cardy\cite{Cardy}
that there might be a four-dimensional
generalisation of  Zamolodchikov's $c$-theorem\cite{Zam}
in two dimensions, such that
there is a function $a(g)$ which has monotonic behaviour under
renormalisation-group (RG) flow
(the strong $a$-theorem) or which
is defined at fixed points such that $a_{\rm UV}-a_{\rm IR}>0$ (the weak
$a$-theorem). It soon became clear that the coefficient of the Gauss-Bonnet
term in the trace of the energy-momentum tensor is the only natural candidate
for the $a$-function in four dimensions. A proof of the weak $a$-theorem has been presented by
Komargodski and Schwimmer\cite{KS} and further analysed and extended in
Refs.\cite{Luty,ElvangST}.

In other work, a perturbative version of the strong $a$-theorem has been
derived\cite{Analog, OsbJacnew}
from Wess-Zumino consistency conditions for the response of
the theory defined on curved spacetime, and with $x$-dependent couplings
$g^I(x)$, to a Weyl rescaling of the metric\cite{Weyl} 
(see Ref.~\cite{OsbJacnew} for a comprehensive set of references).
\footnote{This approach has been extended to six dimensions in 
Refs.\cite{GrinsteinCKA}, and other relevant work on six dimensions appears in 
Refs..~\cite{Fei:2014yja}.} The essential result is that
we can define a function $A$ which satisfies the crucial equation
\be \pa_I A =T_{IJ}\beta^J\, ,
\label{grad}
\ee
 for a function  $T_{IJ}$ which may  in principle be computed perturbatively within the theory extended to curved spacetime and $x$-dependent $g^I$. Eq.~(\ref{grad})
implies
\be
\mu \frac{d}{d\mu} A=\beta^I\frac{\pa}{\pa g^I} A=G_{IJ} \beta^I\beta^J 
\ee
where $G_{IJ}=T_{(IJ)}$; thus verifying the strong $a$-theorem so long as $G_{IJ}$ is positive-definite.  (We shall use the notation $A$ rather than $a$ in anticipation of generalising this equation to three dimensions.)

In odd dimensions there is no Weyl anomaly involving curvature invariants in the usual fashion; though it
does appear in the case of a CP-violating theory with $x$-dependent couplings\cite{Nakthree}.
Therefore it is not
obvious  that this approach to the $a$-function is viable (recall that
in $d=2$ and $d=4$  the natural candidates for the $a$-function  are
associated with  Weyl anomaly terms of the generic form $R$ and $R^2$
(the Gauss-Bonnet term)  respectively). However, the general local RG
approach has been used in three dimensions in Ref.~\cite{Nakthree}
to obtain other consistency
conditions amongst RG quantities. Moreover, it has been
proposed that the free energy in three dimensions  may have similar 
properties to the four-dimensional $a$-function, leading to a conjectured
``$F$-theorem''\cite{jaff,klebb,kleba}. It has been shown that for 
certain theories in three  dimensions, the free energy does indeed
decrease monotonically along RG trajectories. It has also been argued on
general grounds that the  $\beta$-functions should obey a gradient flow
equation in the neighbourhood  of conformal fixed points, with a metric
in Eq.~\eqref{grad} equal to the unit matrix to lowest order. 

In this paper we take a different approach by explicitly constructing a
three-dimensional ``$A$-function''.  We consider a range of renormalisable
theories in three dimensions for which the $\beta$-functions have been
computed at lowest (two-loop) order. This includes general
non-supersymmetric abelian and $SU(n)$ Chern-Simons theories, together
with the supersymmetric $SO(n)$ Chern-Simons theory (the examples of 
supersymmetric $SU(n)$ and $Sp(n)$ being somewhat trivial). Using the
$\beta$-functions,  we show by explicit construction the existence of a
function $A$ satisfying Eq.~\eqref{grad} for a metric which is at
lowest order (and hence perturbatively) positive definite. This is of course the method employed in the classic work of Ref.~\cite{Wallace}. The exact relationship of our $A$-function  
with the $F$-function of Ref.~\cite{jaff,klebb,kleba} is unclear. Certainly an important new feature of our results is that they appear to pass some important higher-order (four-loop) checks. In order to elucidate this further we consider a more general (but ungauged) theory where the relation of the $A$-function with the $\beta$-functions is more transparent.
 
\section{Two-loop results}
We start with the abelian Chern-Simons theory with Lagrangian\cite{kaza}
\begin{align}
L={}&\tfrac12\epsilon^{\mu\nu\rho}A_{\mu}\pa_{\nu}A_{\rho}+|D_{\mu}\phi_j|^2
+i\psibar_j{\hat D}\psi_j\nn
&+\alpha\psibar_j\psi_j\phi^*_k\phi_k+\beta\psibar_j\psi_k\phi^*_k\phi_j
+\tfrac14\gamma (\psibar_j\psi^*_k\phi_j\phi_k
+\psibar^*_j\psi_k\phi^*_j\phi^*_k)-h(\phi^*_j\phi_j)^3.
\label{laga}
\end{align}
where $D_{\mu}=\pa_{\mu}-ig A_{\mu}$ and ${\hat D} = \gamma^{\mu}D_{\mu}$.
 This theory contains equal numbers of scalar-fermion pairs
$(\phi_i,\psi_i), i=1\cdots n$ with the same charge $g$, and  has a
global $SU(n)$ symmetry with respect to which $(\phi_i,\psi_i)$
transform according to the fundamental  representation. It is
renormalisable, and contains five dimensionless couplings,
$\{\alpha,\beta,\gamma, h,g\}$. Of course in Chern-Simons theories 
the gauge coupling $g$ 
has a special 
role as a topological coupling and does not run; its 
$\beta$-function is zero and will not be further mentioned.

The two-loop $\beta$-functions were computed in 
Ref.~\cite{kaza} and were given as
\begin{align}
\beta^{(2)}_{\alpha}
={}&\left(\tfrac83n+2\right)\alpha^3+\tfrac{16}{3}\alpha^2\beta
+\left(\tfrac83n+3\right)\alpha\beta^2+(n+2)\beta^3+
\tfrac14\left(\tfrac83n+\tfrac{17}{3}\right)\alpha\gamma^2\nn
&+\tfrac34(n+2)\beta\gamma^2+3\beta^2g^2+\tfrac14\gamma^2 g^2
-\tfrac23(20n+31)\alpha g^4-8\beta g^4-8(n+2)g^6,\nn
\beta^{(2)}_{\beta}={}&\left(\tfrac83n+6\right)\alpha^2\beta
+\left(3n+\tfrac{16}{3}\right)\alpha\beta^2+\left(\tfrac23n+1\right)\beta^3
+\tfrac34(n+2)\alpha\gamma^2\nn
&+\tfrac14\left(\tfrac83n+\tfrac{17}{3}\right)\beta\gamma^2-3n\beta^2g^2
+\tfrac14(n+2)\gamma^2g^2-\tfrac23(8n+31)\beta g^4,\nn
\beta^{(2)}_{\gamma}={}&\left(\tfrac83n+6\right)\alpha^2\gamma+
\left(6n+\tfrac{34}{3}\right)\alpha\beta\gamma
+\left(\tfrac83n+6\right)\beta^2\gamma+\tfrac16(n+1)\gamma^3\nn
&+4\alpha\gamma g^2+2(n+1)\beta\gamma g^2-\tfrac23(2n-5)\gamma g^4,\nn
\beta_h^{(2)}={}&12(3n+11)h^2+4h[4n\alpha^2+8\alpha\beta+(n+3)\beta^2]\nn
&+(n+4)h\gamma^2-4(5n+16)hg^4\nn
&-[4n\alpha^4+16\alpha^3\beta+4(n+5)\alpha^2\beta^2+4(n+3)\alpha\beta^3
+(n+3)\beta^4]\nn
&-[(n+6)\alpha^2+(3n+11)(\alpha\beta+\tfrac12\beta^2)]\gamma^2-\tfrac{1}{16}(n+3)\gamma^4
-2(\alpha+\beta)\gamma^2g^2\nn
&+4(n\alpha^2+2\alpha\beta+\beta^2)g^4-\gamma^2g^4+8(n\alpha+\beta)g^6
+4(2n+7)g^8.
\label{betsa}
\end{align}
Here and elsewhere, we suppress a  factor of $(8\pi)^{-1}$ for each loop
order.  These two-loop $\beta$-functions are straightforward to
calculate  using dimensional regularisation (DREG) with minimal
subtraction;  the relevant Feynman graphs have only simple poles in
$\epsilon = 3 - d$, associated of course  with the absence of
one-loop divergences.  

It is straightforward to show that the Yukawa $\beta$-functions satisfy a relation of the form Eq.~\eqref{grad} :
\be
\begin{pmatrix}\pa_{\alpha}A\\ \pa_{\beta}A\\ \pa_{\gamma}A
\end{pmatrix}
=\begin{pmatrix}n&1&0\\1&n&0\\0&0&\tfrac14(n+1)
\end{pmatrix}
\begin{pmatrix}\beta^{(2)}_{\alpha}\\ \beta^{(2)}_{\beta}\\ 
\beta^{(2)}_{\gamma} 
\label{yukmet}
\end{pmatrix},
\ee
where
\begin{align}
A={}&\tfrac n4\left(\tfrac83n+2\right)\alpha^4+
\tfrac 16\left(n^2+3n+3\right)\beta^4
+\tfrac{1}{96}(n+1)^2\gamma^4+\left(\tfrac83n+2\right)\alpha^3\beta\nn
&+\tfrac13(3n^2+8n+3)\beta^3\alpha
+\tfrac13(4n^2+9n+8)\alpha^2\beta^2\nn
&+\tfrac{1}{12}(4n+9)(n+1)(\alpha^2+\beta^2)\gamma^2
+\tfrac{1}{12}(9n+17)(n+1)\alpha\beta\gamma^2\nn
&+(1-n^2)\beta^3g^2+\tfrac12(n+1)\alpha\gamma^2g^2
+\tfrac14(n+1)^2\beta\gamma^2g^2\nn
&-\tfrac n3(20n+31)\alpha^2g^4-\tfrac 13(8n^2+31n+12)\beta^2g^4\nn
&-\tfrac n3(2n-5)\gamma^2g^4-\tfrac23(20n+31)\alpha\beta g^4
-8n(n+2)\alpha g^6.
\label{Aabel}
\end{align}
The ``metric'' here is clearly positive-definite (except for the special case $n=1$ where there is a zero eigenvalue reflecting
the equivalence of the $\alpha$, $\beta$ couplings in this case). Of course the metric and $A$-function defined by 
Eq.~\eqref{grad} are only defined up to an overall scale, in the absence of any known relation to other RG quantities such as Weyl anomaly coefficients.

This is on the face of it a remarkable result, and we shall now show that it extends to the other 
three-dimensional theories considered explicitly in Refs.~\cite{kaza,kazb}. We turn first to the non-abelian $SU(n)$ Chern-Simons theory with Lagrangian
\begin{align}
L={}&\tfrac12\epsilon^{\mu\nu\rho}A^A_{\mu}\pa_{\nu}A^A_{\rho}
+\tfrac16g f^{ABC}\epsilon^{\mu\nu\rho}A_{\mu}^AA_{\nu}^BA_{\rho}^C
+|\Dcal_{\mu}\phi_j|^2
+i\psibar_j{\hat \Dcal}\psi_j\nn
&+\alpha\psibar_j\psi_j\phi^*_k\phi_k+\beta\psibar_j\psi_k\phi^*_k\phi_j
+\tfrac14\gamma(\psibar_j\psi^*_k\phi_j\phi_k
+\psibar^*_j\psi_k\phi^*_j\phi^*_k)-h(\phi^*_j\phi_j)^3,
\label{lagb}
\end{align}
where
\be
\Dcal_{\mu}\phi_j=\pa_{\mu}\phi_j-ig T^A_{jk}A^A_{\mu}\phi_k,
\ee
 (similarly for $\Dcal_{\mu}\psi$) with $T^A_{jk}$ the generators for the fundamental  representation of $SU(n)$,
 satisfying $[T^A,T^B]=if^{ABC}T^C$.
The non-gauge interaction terms for this theory are identical with those in
the abelian theory of Eq.~\eqref{laga}. 
The two-loop $\beta$-functions for this theory were computed in 
Ref.~\cite{kazb}; we quote here the Yukawa $\beta$-functions which are given by
\begin{align}
\beta^{(2)}_{\alpha}
={}&\left(\tfrac83n+2\right)\alpha^3+\tfrac{16}{3}\alpha^2\beta
+\left(\tfrac83n+3\right)\alpha\beta^2+(n+2)\beta^3+
\tfrac14\left(\tfrac83n+\tfrac{17}{3}\right)\alpha\gamma^2\nn
&+\tfrac34(n+2)\beta\gamma^2-\alpha\beta g^2+\frac{n^2-3}{2n}\beta^2g^2
+\frac{n^2-1}{8n}\gamma^2g^2\nn
&-\frac{40n^3-17n^2-40n+62}{12n^2}\alpha g^4
-\frac{5n^3+6n^2-18n+8}{4n^2}\beta g^4\nn
&+\frac{3n^4-4n^3+5n^2-8n+16}{8n^3}g^6,\nn
\beta^{(2)}_{\beta}={}&\left(\tfrac83n+6\right)\alpha^2\beta
+\left(3n+\tfrac{16}{3}\right)\alpha\beta^2+\left(\tfrac23n+1\right)\beta^3
+\tfrac34(n+2)\alpha\gamma^2\nn
&+\tfrac14\left(\tfrac83n+\tfrac{17}{3}\right)\beta\gamma^2+n\alpha\beta g^2+\beta^2g^2
+\frac{n^2-1}{4n}\gamma^2 g^2-\frac{5(n^2-4)}{4n}\alpha g^4\nn
&-\frac{22n^3-23n^2-64n+62}{12n^2}\beta g^4
-\frac{(n^2-4)(n-2)}{2n^2}g^6,\nn
\beta^{(2)}_{\gamma}={}&\left(\tfrac83n+6\right)\alpha^2\gamma+
\left(6n+\tfrac{34}{3}\right)\alpha\beta\gamma
+\left(\tfrac83n+6\right)\beta^2\gamma+\tfrac16(n+1)\gamma^3\nn
&+\frac{(n-1)(n+2)}{n}\alpha\gamma g^2+\frac{(n-1)(2n+1)}{n}\beta\gamma g^2\nn
&-\frac{(n-1)(2n^2-2n+5)}{6n^2}\gamma g^4.
\label{betsb}
\end{align}
The non-gauge terms in Eq.~\eqref{betsb} are of course identical with those in 
Eq.~\eqref{betsa}.

The corresponding $A$-function (with a metric identical to 
that appearing in the abelian case, Eq.~\eqref{yukmet}), is given by
\begin{align}
A={}&\tfrac n4\left(\tfrac83n+2\right)\alpha^4+
\tfrac 16\left(n^2+3n+3\right)\beta^4
+\tfrac{1}{96}(n+1)^2\gamma^4+\left(\tfrac83n+2\right)\alpha^3\beta\nn
&+\tfrac13(3n^2+8n+3)\beta^3\alpha
+\tfrac13(4n^2+9n+8)\alpha^2\beta^2\nn
&+\tfrac{1}{12}(4n+9)(n+1)(\alpha^2+\beta^2)\gamma^2
+\tfrac{1}{12}(9n+17)(n+1)\alpha\beta\gamma^2\nn
&+(n^2-1)\Bigl[\frac{n+2}{8n}\alpha\gamma^2g^2
+\frac{2n+1}{8n}\beta\gamma^2g^2+\tfrac12\alpha\beta^2g^2
+\frac{1}{2n}\beta^3g^2\nn
&-\frac{20n-1}{12n}\alpha^2g^4-\frac{11n^2-4n-12}{12n^2}
\beta^2g^4-\frac{2n^2-2n+5}{48n^2}\gamma^2g^4\nn
&-\frac{15n^2+40n-62}{12n^2}\alpha\beta g^4
+\frac{3n^2-8n+16}{8n^2}\alpha g^6-\frac{4n^3-11n^2-8n+16}{8n^3}\beta g^6
\Bigr].
\end{align}
Once again, the non-gauge terms are identical to those in Eq.~\eqref{Aabel}.
 
We have so far not discussed the scalar $\beta$-function $\beta_h$. 
For this we shall return to the abelian case of Eq.~\eqref{laga}.
We expect a relation of the form 
\be
\pa_hA=X\beta_h^{(2)}+\ldots
\label{hderiv}
\ee
where in principle the right-hand side could involve the Yukawa 
$\beta$-functions as well. However, it turns out that no such additional
terms are required at this order. If we retain only $\beta_h$ in
Eq.~\ref{hderiv} and integrate using Eq.~\eqref{betsa} (assuming
$X$ is independent of $h$), we find that we need to modify $A$ in Eq.~\eqref{Aabel}
according to $A\rightarrow A+A_h$ where 
\begin{align}
A_h={}&X\Bigl[4(3n+11)h^3+2h^2[4n\alpha^2+8\alpha\beta+(n+3)\beta^2]\nn
&+\tfrac12(n+4)h^2\gamma^2-2(5n+16)h^2g^4\nn
&-h[4n\alpha^4+16\alpha^3\beta+4(n+5)\alpha^2\beta^2+4(n+3)\alpha\beta^3
+(n+3)\beta^4]\nn
&-h[(n+6)\alpha^2+(3n+11)(\alpha\beta+\tfrac12\beta^2)]\gamma^2
-\tfrac{1}{16}(n+3)h\gamma^4
-2h(\alpha+\beta)\gamma^2g^2\nn
&+4h(n\alpha^2+2\alpha\beta+\beta^2)g^4-h\gamma^2g^4+8h(n\alpha+\beta)g^6
+4h(2n+7)g^8.
\label{Ah}
\end{align}

It is then clear that we need to consider higher-order contributions to Eq.~\eqref{yukmet},
since Eq.~\eqref{Ah} will yield contributions to $\pa_{\alpha} A$ of the form 
$\alpha h^2$ and $\alpha^3h$ which are produced by four-loop diagrams\footnote{This is analogous to the ``3-2-1'' phenomenon discussed in Ref.~\cite{sann}.}. At the very least we
should include four-loop contributions to $\beta_{\alpha}$, etc on the right-hand side of Eq.~\eqref{yukmet}.
 We have accordingly 
computed the four-loop contributions to $\beta$-functions for $\alpha$, 
$\beta$, $\gamma$ with one Yukawa coupling and two scalar couplings; and with three Yukawa couplings and one scalar coupling. 
The results are
\begin{align}
\beta_{\alpha}^{(4)}={}&h^2[\tfrac83(n+1)(n+2)\alpha+2(n+2)\beta]\nn
&-\tfrac23(n+2)h\Bigl\{4(n+1)\alpha^3
+10(n+2)\alpha^2\beta
+(2n+9)\alpha\beta^2+(n+3)\beta^3\nn
&+\tfrac14[(2n+11)\alpha+(3n+11)\beta]\gamma^2\Bigr\}+\ldots,\nn
\beta_{\beta}^{(4)}={}&\tfrac23(n+2)(n+4)h^2\beta\nn
&-\tfrac23(n+2)h\Bigl\{2(n+6)\alpha^2\beta+(3n+10)
\alpha\beta^2+(n+3)\beta^3\nn
&+\tfrac14[3(n+4)\alpha+(3n+11)\beta]\gamma^2\Bigr\}+\ldots,\nn
\beta_{\gamma}^{(4)}={}&\tfrac23(n+2)(n+4)h^2\gamma
-\tfrac43(n+2)h\gamma\left[(n+6)\alpha^2+(3n+11)(\alpha\beta+\tfrac12\beta^2)\right]+\ldots,
\label{hbets}
\end{align}
where the ellipses indicate pure Yukawa contributions which we have not calculated. 
In general, the metric which we have given to leading order in
Eq.~\eqref{yukmet} might be expected to have additional higher-order 
corrections. However, it turns out that as far as the terms in $A$ 
in Eq.~\eqref{Ah} are concerned, no corrections are required and provided
we take
\be
X=\tfrac16(n+1)(n+2)
\ee
we have
\begin{align}
\pa_{\alpha} A_h={}&n\beta_{\alpha}^{(4)}+\beta_{\beta}^{(4)},\nn
\pa_{\beta} A_h={}&\beta_{\alpha}^{(4)}+n\beta_{\beta}^{(4)},\nn
\pa_{\gamma} A_h={}&\tfrac14(n+1)\beta_{\gamma}^{(4)}.
\label{hafun}
\end{align}
 We shall see later that we can  shed more light on the situation by turning to the general (but non-gauged) case; but for the present we continue in the 
next section by considering the supersymmetric gauge theory.
\section{The supersymmetric case}

We turn to the non-abelian $\Ncal=1$ supersymmetric case. Here we consider 
an action
\begin{align}
S={}&\int d^3xd^2\theta\Bigl[-\tfrac14(D^{\alpha}\Gamma^{A\beta})
(D_{\beta}\Gamma_{\alpha}^A)-\tfrac16gf^{ABC}(D^{\alpha}\Gamma^{A\beta})
\Gamma_{\alpha}^B\Gamma_{\beta}^C-\tfrac{1}{24}g^2f^{ABC}f^{ADE}
\Gamma^{B\alpha}\Gamma^{C\beta}\Gamma_{\alpha}^D\Gamma_{\beta}^E\nn
&-\tfrac12(D^{\alpha}\Phibar_j+ig\Phibar_kT^A_{kj}\Gamma^{\alpha}_A)
(D_{\alpha}\Phi_j-ig\Gamma_{\alpha}^BT^B_{jl}\Phi_l)\nn
&+\tfrac14\eta_0(\Phibar_j\Phi_j)^2
+\tfrac14\eta_1(\Phibar_jT^A_{jk}\Phi_k)^2\Bigr],
\label{lagsusy}
\end{align}
where $\Gamma^{A\alpha}$ is a real gauge superfield and $\Phi$ a complex matter supermultiplet (once again in the fundamental representation). Note that here $\alpha$ and $\beta$ are three-dimensional spinor indices. We have a gauge coupling $g$ and matter couplings $\eta_0$, $\eta_1$. 
We refer the reader to Refs.~\cite{kaza, kazb} for the definition of the supercovariant derivative $D_{\alpha}$ and other details of the notation and conventions. Of course in the abelian and also the non-abelian $SU(n)$ case, this action
is already included in the general cases of Eqs.~\eqref{laga}, \eqref{lagb}.

The $\beta$-functions are given by
\footnote{Since we now deal with supersymmetry, the issue of use 
of regularisation by dimensional reduction (DRED)~\cite{Siegel:1979wq}\ 
rather than DREG arises; 
but since as indicated earlier, the relevant two-loop graphs have simple poles 
in $\epsilon$, minimal subtraction gives the same result in both cases.}
\begin{align}
\beta^{(2)}_{\eta_1}={}&\bigl[(R_{31}+\tfrac12R_{t1}+T_RC_R+2C_R^2)\eta_1
+\tfrac14T_RC_Ag^2-\tfrac12R_{f1}(\eta_1+g^2)\nn
&-\tfrac14C_RC_A(5\eta_1-3g^2)
+\tfrac18C_A^2(\eta_1-3g^2)\bigr](\eta_1^2-g^4)\nn
&+[T_R(\eta_1^2+\eta_1g^2+g^4)+C_R(3\eta_1^2+4\eta_1g^2+3g^4)\nn
&-\tfrac14C_A(5\eta_1^2+8\eta_1g^2+7g^4)]R_{21}(\eta_1-g^2)\nn
&+\bigl[(6R_{21}+10C_R+3T_R-\tfrac32C_A)\eta_1^2+(2n+11)\eta_1\eta_0\nn
&+2C_R\eta_1g^2-(2R_{21}+\tfrac12C_A)g^4\bigr]\eta_0,\nn
\beta^{(2)}_{\eta_0}={}&\bigl[(R_{30}+\tfrac12R_{t0})\eta_1
-\tfrac12R_{f0}(\eta_1+g^2)\bigl](\eta_1^2-g^4)\nn
&+[T_R(\eta_1^2+\eta_1g^2+g^4)+C_R(3\eta_1^2+4\eta_1g^2+3g^4)\nn
&-\tfrac14C_A(5\eta_1^2+8\eta_1g^2+7g^4)]R_{20}(\eta_1-g^2)\nn
&+[7C_R\eta_1\eta_0+3(n+2)\eta_0^2+C_R\eta_0g^2\nn
&+2R_{20}(3\eta_1^2-g^4)+(2C_R+2T_R-C_A)C_R(\eta_1^2-g^4)]\eta_0.
\label{susbets}
\end{align}
Here the ``reduction'' coefficients $R_{Xi}$ 
(where $X$ takes the values $2,3,t,f$ and $i=0,1$) are defined by 
\be
T_X=R_{X0}T_0+R_{X1}T_1,
\ee
where
\begin{align}
T_0=(\Phibar\Phi)^2,&\quad T_1=(\Phibar T^A\Phi)^2,\nn 
T_2=(\Phibar T^AT^B\Phi)^2,&\quad T_3=(\Phibar T^AT^BT^C\Phi)^2,\nn
T_t=(\Phibar T^A\Phi)(\Phibar T^BT^C\Phi)\tr(T^A\{T^B,T^C\}),&\quad
T_f=f^{EAC}f^{EDB}(\Phibar T^AT^B\Phi)(\Phibar T^CT^D\Phi).
\end{align}
The group invariants are defined as usual by
\be
C_R 1=T^AT^A,\quad T_R\delta^{AB}=\tr(T^AT^B),\quad C_A\delta^{AB}=f^{ACD}f^{BCD}.
\ee
As shown in Ref.~\cite{kazb}, the reduction coefficients may be computed for the 
classical groups $SU(n)$, $SO(n)$ and $Sp(n)$; however for  $SU(n)$ the identity
\be
T^A_{ij}T^A_{kl}=T_R(\delta_{il}\delta_{kj}-\tfrac1n\delta_{ij}\delta_{kl})
\label{groupid}
\ee
yields a simple relation between $T_0$ and $T_1$, and a similar argument (with a different identity) holds for $Sp(n)$;
so that effectively 
$\eta_{0,1}$ may be replaced by a single coupling. This is not the case for 
$SO(n)$, where the analogous identity to Eq.~\eqref{groupid} does not reduce the number of quartic scalar invariants. In this case there is therefore potentially a non-trivial result for the $A$-function, so we confine our attention to this gauge group for the remainder of this section. 
We have for $SO(n)$\cite{kazb}
\begin{align}
C_R=\tfrac12(n-1)T_R,&\quad C_A=(n-2)T_R   \nn
R_{20}=\tfrac14(n-1)T_R^2,\quad R_{30}=R_{f0}={}&-\tfrac18T_R^3(n-1)(n-2),\quad
R_{t0}=R_{t1}=0,\nn
R_{21}=-\tfrac12T_R(n-2),\quad R_{31}={}&\tfrac14T_R^2(n^2-3n+3),\quad
R_{f1}=\tfrac14T_R^2(n-2)(n-3),
\end{align}
so that the $\beta$-functions of Eq.~\eqref{susbets} take the form
\begin{align}
\beta^{(2)}_{\eta_0}={}&(n-1)\bigl[\tfrac18T_R^3(5\eta_1^2+6\eta_1g^2+5g^4)(\eta_1-g^2)
+T_R^2(3\eta_1^2-2g^4)\eta_0\nn
&+\tfrac12T_R(7\eta_1+g^2)\eta_0^2\bigr]+3(n+2)\eta_0^3,\nn
\beta^{(2)}_{\eta_1}={}&T_R^2\bigl[\tfrac54\eta_1^3+\tfrac12(n-2)\eta_1^2g^2
+\tfrac14(3-4n)\eta_1g^4+\tfrac12(n-2)g^6\bigr]\nn
&+\bigl[\tfrac12(n+14)\eta_1^2+(n-1)\eta_1g^2+\tfrac12(n-2)g^4\bigr]T_R\eta_0
+(2n+11)\eta_1\eta_0^2.
\end{align}
The value of $T_R$ depends on the choice of scale for the representation matrices and structure constants.There are several possible convenient choices in the $SO(n)$ case; see Ref.~\cite{Lum} for a discussion.
It is straightforward to show that the $\beta$-functions satisfy
\be
\begin{pmatrix}\pa_{\eta_0}A\\ \pa_{\eta_1}A
\end{pmatrix}
=\begin{pmatrix}\frac{4(n+1)}{(n-1)}&2T_R\\2T_R&3T_R^2
\end{pmatrix}
\begin{pmatrix}\beta_{\eta_0}\\ \beta_{\eta_1} 
\end{pmatrix},
\label{susya}
\ee
where
\begin{align}
A={}&\frac{3(n+1)(n+2)}{n-1}\eta_0^4
+\tfrac23(n+1)T_Rg^2\eta_0^3-\tfrac12(7n+10)T_R^2g^4\eta_0^2\nn
&-\tfrac32(n+3)T_R^3g^6\eta_0+6(n+2)T_R\eta_0^3\eta_1+\tfrac{13}{2}(n+2)
T_R^2\eta_0^2\eta_1^2\nn
&+(n-1)T_R^2g^2\eta_0^2\eta_1-\tfrac12(5n-2)T_R^3g^4\eta_0\eta_1
+\tfrac32(n-1)T_R^3g^2\eta_0\eta_1^2\nn
&+\tfrac52(n+2)T_R^3\eta_0\eta_1^3+\tfrac{5}{16}(n+2)T_R^4\eta_1^4
+\tfrac{1}{12}(7n-13)T_R^4g^2\eta_1^3\nn
&-\tfrac18(13n-10)T_R^4g^4\eta_1^2 +\tfrac14(n-7)T_R^4g^6\eta_1.
\label{susyb}
\end{align}
We see from Eq.~\eqref{susya} that the metric is again positive definite.

In Eqs.~\eqref{susya}, \eqref{susyb} we have extended the range of our results to include the $\Ncal=1$
supersymmetric $SO(n)$ theory, in addition to the general abelian and $SU(n)$
theories considered earlier. There is little to be gained from detailed consideration of
the $Sp(n)$ supersymmetric theory; since there is in this case
effectively only a single coupling, the existence of an $A$-function
satisfying Eq.~\eqref{grad} is trivially guaranteed.

Finally a word is in order concerning the relation between the general $SU(n)$ theory of Section 2 and the supersymmetric 
$SU(n)$ theory defined in this section. One may obtain the supersymmetric $SU(n)$ theory of Eq.~\eqref{lagsusy} from the general lagrangian of 
Eq.~\eqref{lagb} by imposing\cite{kazb}
\begin{align}
g\to\frac{g}{\sqrt2},\quad \alpha\to{}&\frac{1}{4n}[(n-1)\eta_0+ng^2],\quad \beta\to\frac{1}{4n}[(n-1)\eta_0-g^2],\nn
\gamma\to\frac{n-1}{2n}(\eta_0-g^2),\quad & h\to\frac{\eta_0^2(n-1)^2}{16n^2},
\end{align}
where we have set  $\eta_1=0$ since again for $SU(n)$ there is effectively only one coupling. It is interesting that due to the replacement $h\sim\eta_0^2$, $A_h$  in Eq.~\eqref{Ah} will be entirely $O(\eta_0^6)$. The terms in $h$ which in the general case would have yielded the two-loop $\beta_h$ (upon differentiation with respect to $h$) will now give rise (upon differentiation with respect to $\eta_0$) to higher-order contributions to the metric in Eq.~\eqref{grad}  multiplying the two-loop $\beta_{\eta_0}$. The connection between the four-loop and two-loop $\beta$-function coefficients  is no longer apparent. A full four-loop analysis (in which it would be 
appropriate to use DRED rather than DREG) might reveal equivalent constraints 
on the $\beta$-function; or alternatively the requisite 
relations between $\beta$-function coefficients might be an 
automatic consequence of supersymmetry. 

\section{General theory}
The above results present, at first sight, persuasive evidence for
an $a$-theorem. It is natural to consider a generalisation 
to a theory with arbitrary scalar and fermion content.
Thus we consider the theory
 
\be
L=\tfrac12[\epsilon^{\mu\nu\rho}A_{\mu}\pa_{\nu}A_{\rho}+
(D_{\mu}\phi_i)^2+i\psibar_a{\hat D}\psi_a]+\tfrac14 Y_{abij}\psi_a\psi_b\phi_i\phi_j
+\tfrac{1}{6!}h_{ijklmn}\phi_i\phi_j\phi_k\phi_l\phi_m\phi_n
\label{lagc} 
\ee
where  we employ a real basis for both scalar and fermion fields.
(Recall that in $d=3$, $\psibar = \psi^{*T}$, and there is  no obstacle to
decomposing $\psi$  into real Majorana fields.)

For the purposes of this paper we shall restrict ourselves to considering the general theory in the ungauged case. 
The result for the two-loop Yukawa $\beta$-function may then be written as 
\be
\beta^{(2)}_{abij}=a\beta_{abij}^{(a)}+b\beta_{abij}^{(b)}+c\beta_{abij}^{(c)}
+d\beta_{abij}^{(d)}+e\beta_{abij}^{(e)},
\ee
where $\beta_{abij}^{(a-e)}$ correspond to the tensor structures  given by
\begin{align}
\beta_{abij}^{(a)}={}&Y_{acil}Y_{cdjm}Y_{dblm}+Y_{aclm}Y_{cdjm}Y_{dbil}+Y_{acjl}Y_{cdim}Y_{dblm}
+Y_{aclm}Y_{cdim}Y_{dbjl},\nn
\beta_{abij}^{(b)}={}&Y_{aclm}Y_{cdij}Y_{dblm},\nn
\beta_{abij}^{(c)}={}&Y_{cdik}Y_{abkl}Y_{cdlj},\nn
\beta^{(d)}_{abij}={}&Y_{acij}Y_{cdlm}Y_{dblm}+Y_{adlm}Y_{dclm}Y_{cbij},\nn
\beta^{(e)}_{abij}={}&Y_{abik}Y_{cdkl}Y_{dclj}+Y_{cdil}Y_{dclk}Y_{abkj},
\label{genbets}
\end{align}
and where 
\be
a= b=c=2, \quad d=e=\tfrac23.
\label{vals}
\ee
(Note that $\beta^{(d,e)}_{abij}$ correspond to fermion and scalar anomalous dimension contributions.)
Upon specialising to the theory of Eq.~\eqref{laga}, we obtain contributions to the $\beta$-functions of the three 
couplings $\alpha$, $\beta$ and $\gamma$ given for $\beta^{(a)}$  by
\begin{align}
\beta^{(a)}_{\alpha}={}&4\alpha^3+6\alpha\beta^2+2n\beta^3+\tfrac12\gamma^2[3\alpha+(3n+4)\beta]\nn
\beta^{(a)}_{\beta}={}&12\alpha^2\beta+6n\alpha\beta^2+2\beta^3+
\tfrac12\gamma^2[3(n+2)\alpha+(2n+3)\beta]\nn
\beta^{(a)}_{\gamma}={}&4\gamma[3\alpha^2+3(n+1)\alpha\beta+(n+2)\beta^2],
\label{sepa}
\end{align}
for $\beta^{(b,c)}$  by
\begin{align}
\beta^{(b)}_{\alpha}=\beta^{(c)}_{\alpha}={}&2n\alpha^3+4\alpha^2\beta+2n\alpha\beta^2+2\beta^3+
\tfrac12\gamma^2[(n+1)\alpha+\beta],\nn
\beta^{(b)}_{\beta}=\beta^{(c)}_{\beta}={}&2n\alpha^2\beta+4\alpha\beta^2+\tfrac12\beta\gamma^2,\nn
\beta^{(b)}_{\gamma}=\beta^{(c)}_{\gamma}={}&2\gamma(n\alpha^2+2\alpha\beta+\beta^2),
\label{sepb}
\end{align}
and for the anomalous dimension contributions $\beta^{(d,e)}$ by
\begin{align}
\beta^{(d)}_{\alpha}=\beta^{(e)}_{\alpha}={}&\gtil\alpha,\nn
\gtil={}&2(n\alpha^2+2\alpha\beta+n\beta^2)+\tfrac12(n+1)\gamma^2,
\label{sepc}
\end{align}
with similar results for $\beta^{(d,e)}_{\beta,\gamma}$. It is easy to check that upon adding the contributions to
$\beta_{\alpha,\beta,\gamma}$ from Eqs.~\eqref{sepa}, \eqref{sepb}, \eqref{sepc}, and incorporating the factors of
$a$-$e$ from Eq.~\eqref{vals}, we obtain the $\beta$-functions given in Eq.~\eqref{betsa} (upon setting the gauge coupling $g=0$) up to an overall factor of 4 which we are unable to account for. Of course such an overall factor has no effect on the existence or otherwise of the $A$-function.

It can readily be seen that for $\beta$-functions given by Eq.~\eqref{genbets}, the $A$-function given by
\be
A=aA_a+\tfrac14bA_b+\tfrac14cA_c+\tfrac12dA_d+\tfrac12eA_e,
\label{Agen}
\ee
where
\begin{align}
A_a={}&Y_{abij}Y_{bckl}Y_{cdik}Y_{dajl},\nn
A_b={}&Y_{abij}Y_{bckl}Y_{cdij}Y_{dakl},\nn
A_c={}&Y_{abij}Y_{cdjk}Y_{abkl}Y_{cdli},\nn
A_d={}&Y_{acij}Y_{cbij}Y_{bdlm}Y_{dalm},\nn
A_e={}&Y_{abil}Y_{balj}Y_{cdil}Y_{dclj}.
\label{Adefs}
\end{align}
satisfies Eq.~\eqref{grad} in the form
\be
\frac{\pa A}{\pa Y_{abij}}=\beta^{(2)}_{abij},
\label{gradtwo}
\ee
where we define 
\be
\frac{\pa }{\pa Y_{abij}}Y_{a'b'i'j'}=\tfrac14(\delta_{aa'}\delta_{bb'}+\delta_{ab'}\delta_{ba'})
(\delta_{ii'}\delta_{jj'}+\delta_{ij'}\delta_{ji'}).
\ee

The corresponding (lowest order) contribution to the metric $G_{IJ}$ 
is therefore effectively the unit matrix in coupling space. The different tensor structures in Eq.~\eqref{Adefs}
are depicted in Table~\ref{fig1} where a vertex represents a $Y$ and full (``fermion'') and dashed (``scalar'') lines represent contractions of 
$a,b,\ldots$ and $i,j,\ldots$ indices, respectively. Differentiation with respect to $Y$ therefore corresponds to removing a
vertex. and is easily seen to produce a potential $\beta$-function contribution.
\begin{table}[h]
	\setlength{\extrarowheight}{1cm}
	\setlength{\tabcolsep}{24pt}
	\hspace*{-5.75cm}
	\centering
	\resizebox{6.7cm}{!}{
		\begin{tabular*}{20cm}{ccccc}
			\begin{picture}(162,162) (287,-207)
			\SetWidth{1.0}
			\SetColor{Black}
			\Arc(368,-126)(80,127,487)
			\Line[dash,dashsize=10](305,-78)(305,-175)
			\Line[dash,dashsize=9.4](305,-175)(433,-79)
			\Line[dash,dashsize=10](433,-79)(433,-173)
			\Line[dash,dashsize=8.2](433,-173)(305,-78)
			\end{picture}
			&
			\begin{picture}(162,162) (287,-207)
			\SetWidth{1.0}
			\SetColor{Black}
			\Arc(368,-126)(80,127,487)
			\Arc[dash,dashsize=10](493.891,-126.5)(148.892,147.728,212.272)
			\Arc[dash,dashsize=10](245.793,-125.728)(146.213,-33.299,32.578)
			\Arc[dash,dashsize=10,clock](368.272,-249.207)(146.213,123.299,57.422)
			\Arc[dash,dashsize=10,clock](365.917,-1.121)(148.893,-57.004,-121.554)
			\end{picture}
			&
			\begin{picture}(162,162) (287,-207)
			\SetWidth{1.0}
			\SetColor{Black}
			\Arc(368,-126)(80,127,487)
			\Arc[dash,dashsize=10](437.778,-126.5)(105.779,131.274,228.726)
			\Arc[dash,dashsize=10](302.985,-126.088)(103.019,-50.869,50.148)
			\Arc(369,-126)(37.443,146,506)
			\end{picture}
			&
			\begin{picture}(162,162) (287,-207)
			\SetWidth{1.0}
			\SetColor{Black}
			\Arc(368,-126)(80,127,487)
			\Arc[dash,dashsize=10](465.7,-126.5)(148.708,149.935,210.065)
			\Arc[dash,dashsize=10](159.567,-126.5)(192.439,-22.776,22.776)
			\Arc[dash,dashsize=10](540.898,-125.032)(156.898,153.081,207.715)
			\Arc[dash,dashsize=10](273.562,-125.598)(147.439,-29.41,28.609)
			\end{picture}
			&
			\begin{picture}(162,162) (191,-111)
			\SetWidth{1.0}
			\SetColor{Black}
			\Arc[dash,dashsize=8.4,clock](272.8,-30.4)(79.604,37.446,-90.576)
			\Arc[dash,dashsize=8.4,clock](273.109,-29.284)(80.723,-90.787,-206.711)
			\Arc[dash,dashsize=8.4,clock](271.587,-30.94)(81.019,152.077,33.689)
			\Arc(406.898,-31.816)(182.9,155.855,202.565)
			\Arc(106.024,-28.64)(150.995,-29.068,27.465)
			\Arc(465.342,-31.616)(177.344,155.474,203.736)
			\Arc(152.501,-29.459)(167.506,-26.042,25.252)
			\end{picture}
			\\
			{\Huge $A_{a}$}
			&
			{\Huge $A_{b}$}
			&
			{\Huge $A_{c}$}
			&
			{\Huge $A_{d}$}
			&
			{\Huge $A_{e}$}
		\end{tabular*}
	}
	\caption{Contributions to $A$ from Yukawa couplings}
	\label{fig1}	
\end{table}

Upon specialising once again to the theory of Eq.~\eqref{laga}, we obtain 
\begin{align}
A_a={}&n\alpha^4+4\alpha^3\beta+6n\alpha^2\beta^2+2(n^2+1)\alpha\beta^3+n\beta^4\nn
&+\tfrac12(n+1)\gamma^2[3\alpha^2+3(n+1)\alpha\beta+(n+2)\beta^2],\nn
A_{b,c}={}&\tfrac12[n^2\alpha^4+4n\alpha^3\beta+2(n^2+2)\alpha^2\beta^2+4n\alpha\beta^3+\beta^4
+\tfrac12(n+1)\gamma^2(n\alpha^2+2\alpha\beta+\beta^2)],\nn
A_{d,e}={}&\tfrac12[n^2\alpha^4+4n\alpha^3\beta+2(n^2+2)\alpha^2\beta^2+4n\alpha\beta^3+n^2\beta^4\nn
&+\tfrac12(n+1)\gamma^2(n\alpha^2+2\alpha\beta+n\beta^2)];
\end{align}
and again, it may readily be checked that upon inserting these expressions into Eq.~ \eqref{Agen}, we obtain the $A$-function derived in Eq.~\eqref{Aabel} (with, again, $g=0$). So far, there is no constraint on the two-loop coefficients $a$--$e$ in Eq.~\eqref{vals}. This is because the symmetries of the tensor structures appearing in Eq.~\eqref{Agen} imply a one-to-one relation between $A$-function contributions and Yukawa $\beta$-function contributions. This may be seen in Table~\ref{fig1} where in each diagram, the removal of any vertex leads to the same $\beta$-function contribution. The $A$-function can thus be
tailored term-by-term to match any values for the $\beta$-function coefficients. This situation appears likely to be unchanged if we include gauge contributions, although we have not performed the explicit calculations.
 
 However, a crucial difference arises in the case of the scalar coupling $h_{ijklmn}$. The two-loop $\beta$-function for $h$ is given by
 \be
 (\beta_h^{(2)})_{ijklmn}=h_1\beta_{ijklmn}^{(h_1)}+h_2\beta_{ijklmn}^{(h_2)}+ h_3\beta_{ijklmn}^{(h_3)}+h_4\beta_{ijklmn}^{(h_4)}+h_5\beta_{ijklmn}^{(h_5)},
 \ee
 where
 \begin{align}
 \beta_{ijklmn}^{(h_1)}=\tfrac{1}{6!}(h_{ijkpqr}h_{lmnpqr}+\hbox{perms}), \quad &
 \beta_{ijklmn}^{(h_2)}=\tfrac{1}{6!}(h_{ijklpq}Y_{abmp}Y_{abnq}+\hbox{perms}),\nn
 \beta_{ijklmn}^{(h_3)}=\tfrac{1}{6!}(h_{ijklmp}Y_{abpq}Y_{abnq}+\hbox{perms}),\quad&
 \beta_{ijklmn}^{(h_4)}=\tfrac{1}{6!}(Y_{abij}Y_{bckl}Y_{cdmp}Y_{dapn}+\hbox{perms}), \nn
 \beta_{ijklmn}^{(h_5)}={}&\tfrac{1}{6!}(Y_{abij}Y_{bcmp}Y_{cdkl}Y_{dapn}+\hbox{perms}),
 \end{align}
where ``+ perms'' completes the $6!$ permutations of the indices $\{ijklmn\}$ and 
 \be
 h_1=\tfrac{40}{3},\quad h_2=30, \quad h_3=4,\quad h_4=h_5=-360.
 \ee
 Eq.~\eqref{grad} may be satisfied for the coupling $h_{ijklmn}$ by introducing extra $A$-function contributions 
 \be
 A\rightarrow A+h_1A_{h_1}+h_2 A_{h_2}+h_3A_{h_3}+h_4A_{h_4}+h_5A_{h_5},
 \label{Afour}
 \ee
 where
 \begin{align}
 A_{h_1}=\tfrac13\lambda h_{ijklmn}\beta_{ijklmn}^{(h_1)},\quad 
 A_{h_2}={}&\tfrac12\lambda h_{ijklmn}\beta_{ijklmn}^{(h_2)},\quad
  A_{h_3}=\tfrac12\lambda h_{ijklmn}\beta_{ijklmn}^{(h_3)},\nn
  A_{h_4}=\lambda h_{ijklmn}\beta_{ijklmn}^{(h_4)},\quad &
 A_{h_5}=\lambda h_{ijklmn}\beta_{ijklmn}^{(h_5)},
 \label{Afoura}
 \end{align}
 so that manifestly
 \be
\frac{\pa A}{\pa h_{ijklmn}}=\lambda (\beta^{(2)}_h)_{ijklmn}.
\ee
 The different tensor structures in Eq.~\eqref{Afoura}
are depicted in Table~\ref{fig2}. Here differentiation with respect to $h$ corresponds to removing a vertex with six ``scalar'' lines
(an $h$-vertex) while differentiation with respect to $Y$ corresponds as before to removing a $Y$ vertex with two ``scalar'' and two ``fermion'' lines.

\begin{table}[h]
	\setlength{\extrarowheight}{1cm}
	\setlength{\tabcolsep}{24pt}
	\hspace*{-5.75cm}
	\centering
	\resizebox{6.7cm}{!}{
		\begin{tabular*}{20cm}{ccccc}
			\begin{picture}(162,162) (191,-111)
			\SetWidth{1.0}
			\SetColor{Black}
			\Arc[dash,dashsize=10](265.677,85.569)(101.765,-129.461,-43.903)
			\Arc[dash,dashsize=10](433.132,-117.254)(161.295,126.143,177.422)
			\Arc[dash,dashsize=10](113.213,-127.061)(159.701,6.133,57.082)
			\Arc[dash,dashsize=10,clock](272.8,-30.4)(79.604,37.446,-90.576)
			\Arc[dash,dashsize=10,clock](273.109,-29.284)(80.723,-90.787,-206.711)
			\Arc[dash,dashsize=10,clock](271.587,-30.94)(81.019,152.077,33.689)
			\Arc[dash,dashsize=10,clock](277.679,-186.2)(208.425,111.291,72.89)
			\Arc[dash,dashsize=10](363.385,26.038)(163.616,-172.965,-124.377)
			\Arc[dash,dashsize=10,clock](186.907,15.538)(151.101,-0.583,-56.183)
			\end{picture}
			&
			\begin{picture}(162,163) (191,-111)
			\SetWidth{1.0}
			\SetColor{Black}
			\Arc[dash,dashsize=10,clock](272.8,-29.4)(79.604,37.446,-90.576)
			\Arc[dash,dashsize=10,clock](273.109,-28.284)(80.723,-90.787,-206.711)
			\Arc[dash,dashsize=10,clock](271.587,-29.94)(81.019,152.077,33.689)
			\Arc[dash,dashsize=10](306.308,-29.5)(88.309,114.277,245.723)
			\Arc[dash,dashsize=10,clock](244.254,-28.691)(83.746,72.096,-73.528)
			\Arc[dash,dashsize=10](393.926,-29.778)(147.928,146.903,212.381)
			\Arc[dash,dashsize=10,clock](171.143,-29.5)(126.858,38.806,-38.806)
			\Arc(272,-29)(25.612,141,501)
			\end{picture}
			&
			\begin{picture}(162,162) (191,-111)
			\SetWidth{1.0}
			\SetColor{Black}
			\Arc[dash,dashsize=10,clock](272.8,-30.4)(79.604,37.446,-90.576)
			\Arc[dash,dashsize=10,clock](273.109,-29.284)(80.723,-90.787,-206.711)
			\Arc[dash,dashsize=10,clock](271.587,-30.94)(81.019,152.077,33.689)
			\Arc[dash,dashsize=10](313.5,-30.5)(91.501,118.386,241.614)
			\Arc[dash,dashsize=10,clock](238.309,-29.616)(85.692,68.295,-69.728)
			\Arc[dash,dashsize=10](443.614,-31.088)(191.617,154.965,204.319)
			\Arc[dash,dashsize=10,clock](137.364,-30.5)(154.637,30.938,-30.938)
			\Arc(292,-31)(21.471,118,478)
			\end{picture}
			&
			\begin{picture}(162,162) (287,-207)
			\SetWidth{1.0}
			\SetColor{Black}
			\Arc(368,-126)(80,127,487)
			\Arc[dash,dashsize=10](301.579,-129.689)(67.474,2.284,76.789)
			\Arc[dash,dashsize=10](364.609,-76.876)(49.319,164.866,275.108)
			\Arc[dash,dashsize=10,clock](438.809,-130.465)(69.952,176.34,108.166)
			\Arc[dash,dashsize=10,clock](369.646,-77.704)(49.297,16.141,-89.589)
			\Arc[dash,dashsize=9.6](393.489,-215.871)(92.184,105.406,156.423)
			\Arc[dash,dashsize=9.6,clock](345.395,-210.674)(87.903,74.423,21.821)
			\Line[dash,dashsize=9](309,-181)(427,-181)
			\end{picture}
			&
			\begin{picture}(166,180) (285,-207)
			\SetWidth{1.0}
			\SetColor{Black}
			\Bezier[dash,dsize=8.8](369,-28)(441,-76)(292,-140)(369,-188)
			\Arc(368,-108)(80,127,487)
			\Bezier[dash,dsize=8.8](369,-28)(294,-76)(440,-140)(369,-188)
			\Line[dash,dashsize=9.4](289,-108)(448,-108)
			\Arc[dash,dashsize=10](368,-124.347)(81.653,168.451,371.549)
			\end{picture}
			\\
			{\Huge $A_{h_{1}}$}
			&
			{\Huge $A_{h_{2}}$}
			&
			{\Huge $A_{h_{3}}$}
			&
			{\Huge $A_{h_{4}}$}
			&
			{\Huge $A_{h_{5}}$}
		\end{tabular*}
	}
	\caption{Higher order contributions to $A$ from scalar coupling}
	\label{fig2}
\end{table}

 So far there is no constraint on $h_{1-5}$, just as in the case of $a$-$e$. However, when we differentiate $A$ in Eq.~\eqref{Afour} with respect to $Y_{abij}$, we obtain terms with the structure of four-loop contributions to $\beta_{abij}$ (as we saw earlier in Section 2, and easily see by removing $Y$ vertices from diagrams in Table~(\ref{fig2})).  So if Eq.~\eqref{grad} is to remain valid, then the 
 $A$-function  of Eq.~\eqref{Afour} already requires that Eq.~\eqref{gradtwo} should be extended to four-loop order.  In principle at this order Eq.~\eqref{gradtwo} could include higher order contributions to $G_{IJ}$, contracted with the two-loop Yukawa $\beta$-function; however it is easy to see that these do not contribute to the $h$-dependent terms we are considering. We thus obtain from Eqs.~\eqref{gradtwo}, \eqref{Afour} a prediction for
 the $h$-dependent contributions to the {\it four-loop} $\beta$ function for $Y$,  given by
 \begin{align}
 \beta_{abij}^{(4)}={}&\lambda\left[h_2\beta^{(h_2)}_{abij}+\tfrac12h_3\beta^{(h_3)}_{abij}
 +\tfrac12h_4\beta^{(h_4)}_{abij}+h_5\beta^{(h_5)}_{abij}\right]+h\hbox{-independent terms},\nn
\beta_{abij}^{(h_2)}={}&h_{ikmnpq}h_{jlmnpq}Y_{abkl},\nn
\beta_{abij}^{(h_3)}={}&h_{ilmnpq}h_{klmnpq}Y_{abkj}+h_{jlmnpq}h_{klmnpq}Y_{abik},\nn
  \beta^{(h_4)}_{abij}={}&[(Y_{acik}Y_{cdlm}Y_{dbnp}+Y_{acnp}Y_{cdlm}Y_{dbik})h_{jklmnp}\nn
 &+(Y_{ackp}Y_{cdmp}Y_{dbnl}+Y_{acnl}Y_{cdmp}Y_{dbkp})h_{ijklmn}]+(i\leftrightarrow j),\nn
 \beta^{(h_5)}_{abij}={}&[Y_{ackl}Y_{cdjm}Y_{dbnp}h_{iklmnp}+Y_{ackp}Y_{cdlm}Y_{dbnp}h_{ijklmn}]
 +(i\leftrightarrow j).
 \label{expfour}
 \end{align}
Note that in the case of the $h_{4,5}$ diagrams in Table~\ref{fig2}, there are two topologically inequivalent types of $Y$-vertex and therefore differentiation with respect to $Y$ yields two different types of $\beta$-function contribution from each of these diagrams.
Remarkably the predictions  of Eq.~\eqref{expfour} are verified by explicit calculations, provided $\lambda=\tfrac{1}{90}$. This agreement requires relations between two and four loop Feynman diagram pole contributions, and indeed (since as mentioned $h_4$ and $h_5$ each appear in two distinct  $\beta$-function contributions 
in Eq.~\eqref{expfour}) between distinct four-loop Feynman diagrams. Yet again, these results are in agreement with the earlier explicit results, in this case those of Eqs.~\eqref{hbets}, \eqref{hafun}. Since $\lambda\ne1$, the overall metric in $\{Y,h\}$ coupling space is not simply the unit matrix; however, this could of course be arranged by a rescaling of $h$. 
 
\section{Conclusions}
We have demonstrated explicitly the existence of a three-dimensional $A$-function satisfying Eq.~\eqref{grad} for the two-loop $\beta$-functions 
derived earlier in Refs.~\cite{kaza,kazb} for a range of gauge theories coupled to scalars and fermions. Considering a more general (but ungauged) theory, we have shown that at lowest order the existence of the $A$-function is guaranteed irrespective of the values of the coefficients of the two-loop $\beta$-functions for the scalar and Yukawa couplings; but that this $A$-function also entails predictions for terms in the {\it four-loop} Yukawa $\beta$-function. We have verified that these predictions are correct. It would be interesting to investigate the relation between our results and those of Ref.~\cite{kleba}. These authors presented arguments based on conformal field theory for an ``$F$-function'' satisfying Eq.~\eqref{grad} at lowest order, and also found a trivial metric at this order.  However, their calculations did not reveal the subtle interplay between the two and four loop $\beta$ functions which is necessary for consistency in the theory with both fermions and scalars.

It is worth pointing out that these methods could also be employed in four dimensions; indeed,they already were to a considerable extent in Refs.~\cite{OsbJacnew,JacPoole}, though here the metric was known already at leading order from explicit computation. It would appear that in both cases the leading order metric is constant (except for a $\frac{1}{g^2}$ factor in $G_{gg}$ in four dimensions), and owing to the  ``3-2-1'' phenomenon\cite{sann} could be  determined using our present method up to an overall factor.

It is clearly of crucial importance to carry out the full four-loop calculation (at least for the non-gauge theory) to complete the check that Eq.~\eqref{grad} is valid at this level. 
This computation is under way and will be reported on shortly\cite{usnew}. It would also be of interest to extend our lowest order computations to a general scalar-fermion
gauge theory.

Our results certainly appear to point to the all-orders validity of Eq.~\eqref{grad} in three dimensions. One way to confirm this would be to attempt to adapt the renormalisation group approach of Refs.~\cite{Analog}. As commented earlier, this approach is hampered by the lack of natural RG quantities to serve as the natural candidate for an $A$-function. One potentially significant observation is that $R\phi^2$ is dimensionless in all dimensions but there is no clear link between the $\beta$-function for the coefficient of this quantity and the $A$-function as computed here. 

Finally it would be interesting to test the monotonicity of our $A$-function beyond the weak-coupling limit, by examining its behaviour at fixed points. This could in principle be done explicitly; the fixed-point structure of the models considered in Section 2 was already mapped out in Refs.~\cite{kaza, kazb}. More speculatively, one might investigate whether the values of $A$ at fixed points in supersymmetric theories bore any relation to the corresponding values obtained by ``$F$-maximisation''\cite{morita}.

\section{Acknowledgements}

We thank Hugh Osborn for a careful reading of the manuscript and several useful suggestions.
This work was  supported in part by the STFC under contract ST/G00062X/1, 
and CP by a STFC studentship.

\end{document}